\documentclass[aps,pra,twocolumn,floatfix,superscriptaddress]{revtex4-1}
\usepackage{subfigure,dcolumn}
\usepackage[T1]{fontenc}
\usepackage[english]{babel}
\usepackage{braket}
\usepackage{graphicx}
\usepackage[colorinlistoftodos]{todonotes}
\usepackage[utf8]{inputenc}
\usepackage{amsthm}
\usepackage{amsmath}
\usepackage{amsfonts}
\usepackage{slashed}
\usepackage{hyperref}
\usepackage[normalem]{ulem}
\usepackage{soul}
\begin{document}

\title{Single photons by quenching the vacuum}

\author{E. Sánchez-Burillo}
\affiliation{Max-Planck-Institut f\"ur Quantenoptik, D-85748 Garching, Germany}

\author{L. Martín-Moreno}
\affiliation{Instituto de Ciencia de Materiales de Aragón and
  Departamento de Física de la Materia Condensada, CSIC-Universidad de
  Zaragoza, E-50009 Zaragoza, Spain}

\author{J. J. García-Ripoll}
\affiliation{Instituto de Física Fundamental, IFF-CSIC, Calle Serrano
  113b, Madrid E-28006}

\author{D. Zueco}
\affiliation{Instituto de Ciencia de Materiales de Aragón and Departamento de Física de la Materia Condensada, CSIC-Universidad de Zaragoza, E-50009 Zaragoza, Spain}
\affiliation{Fundación ARAID, Paseo María Agustín 36, E-50004 Zaragoza, Spain}

\date{\today}

\begin{abstract}
Heisenberg's uncertainty principle implies that the quantum vacuum is not empty but fluctuates.
These fluctuations can be converted into radiation through nonadiabatic changes in the Hamiltonian.
Here, we discuss how to \emph{control} this vacuum radiation, engineering a single-photon emitter out of a two-level system (2LS) ultrastrongly coupled to a finite-band waveguide in a vacuum state.
More precisely, we show  the 2LS  nonlinearity shapes the vacuum radiation into a nonGaussian superposition of even and odd cat states. When the 2LS bare frequency lays within the band gaps, this emission can be well approximated by individual photons.
This picture is confirmed by a characterization of the ground and bound states, and a study of the dynamics with matrix product states and polaron Hamiltonian methods.
% These are not really analytical.
\end{abstract}

%\pacs{42.50.Ct, 42.50.-p, 03.65.-w, 11.55.Bq}

%42.50.Ex, 42.50.Pq

\maketitle

%%%%%%%%%%%%%%%%%%%%%%%%%%%%%%%%%%%%%%%%%%
%%%%%%%%%%%%%%%%%%%%%%%%%%%%%%%%%%%%%%%%%%
%%%%%%%%%%%%%%%%%%%%%%%%%%%%%%%%%%%%%%%%%%
%%%%%%%%%%%%%%%%%%%%%%%%%%%%%%%%%%%%%%%%%%
\emph {Introduction.-}
%Quantum vacuum fluctuations are a consequence of non-commuting terms in the Hamiltonian.
%as explained within the Heisenberg uncertainty. 
%
%Therefore they are a prominent example in any quantum mechanics course.  
%
Quantum fluctuations underly many physical phenomena, \emph{e.g.} the Lamb Shift \cite{Lamb1947} or a modification of the atomic decay. They also try to explain \cite{Weisskopf1930} the cosmological-constant problem \cite{Wang2017,*Mazzitelli2018,*Wang2018}.
Vacuum fluctuations can be converted into radiation by nonadiabatic changes of the electromagnetic environment \cite{Nation2012}, as in the dynamical Casimir \cite{Casimir1948,Lamoreaux2007,Moore1970,Lhteenmki2013,Wilson2011}, and Unruh effects \cite{Unruh1976}, and the Hawking radiations \cite{Hawking1974,Hawking1975}.
All these processes are explained with free-field theories---quadratic Hamiltonians of harmonic oscillators---that result in Gaussian states \cite{Adesso2014}. 
To create vacuum radiation with nontrivial statistics we need nonlinearities, such as quantum emitters.

In this work we study the conversion of vacuum fluctuations into single-photon radiation. We focus on \emph{waveguide QED} \ \cite{Roy2017}, studying a two-level system (2LS) coupled to a finite-bandwidth environment of one-dimensional bosonic modes. This low-dimensional realization of the spin-boson model\ \cite{Weiss2008} leads to enhanced light-matter interactions. We assume these interactions to be in the ultrastrong coupling regime, where the coupling is comparable to the excitation energy of the quantum emitter\ \cite{Sanchez-Burillo2014, Peropadre2013a,Sanchez-Burillo2015,Gheeraert2017,Forn-Diaz2017,Puertas2018,FriskKockum2019}. Under these conditions, we show how to convert vacuum fluctuations into individual photons.
  Our protocol consists in either abruptly switching on and off the light-matter coupling constant, or
moving the qubit gap in and out the photonic band (we will show the equivalence of both protocols in Supplemental Material).
%\sout{as sketched in Fig. \ref{Fig1}c)}.
%
We demonstrate that this process is mediated by photon bound states, which we characterize numerically and analytically. These states, once the emitter excitation energy approaches the band-gap, allow the emission of individual photons without violating the parity constraints of the model. Finally, we prove that the two-level system serves % not only as a precursor of vacuum radiation but
also as a detector of quantum fluctuations.

%%%%%%%%%%%%%%%%%%%%%%%%%%%%
\begin{figure}[b]
\includegraphics[width=.85\columnwidth]{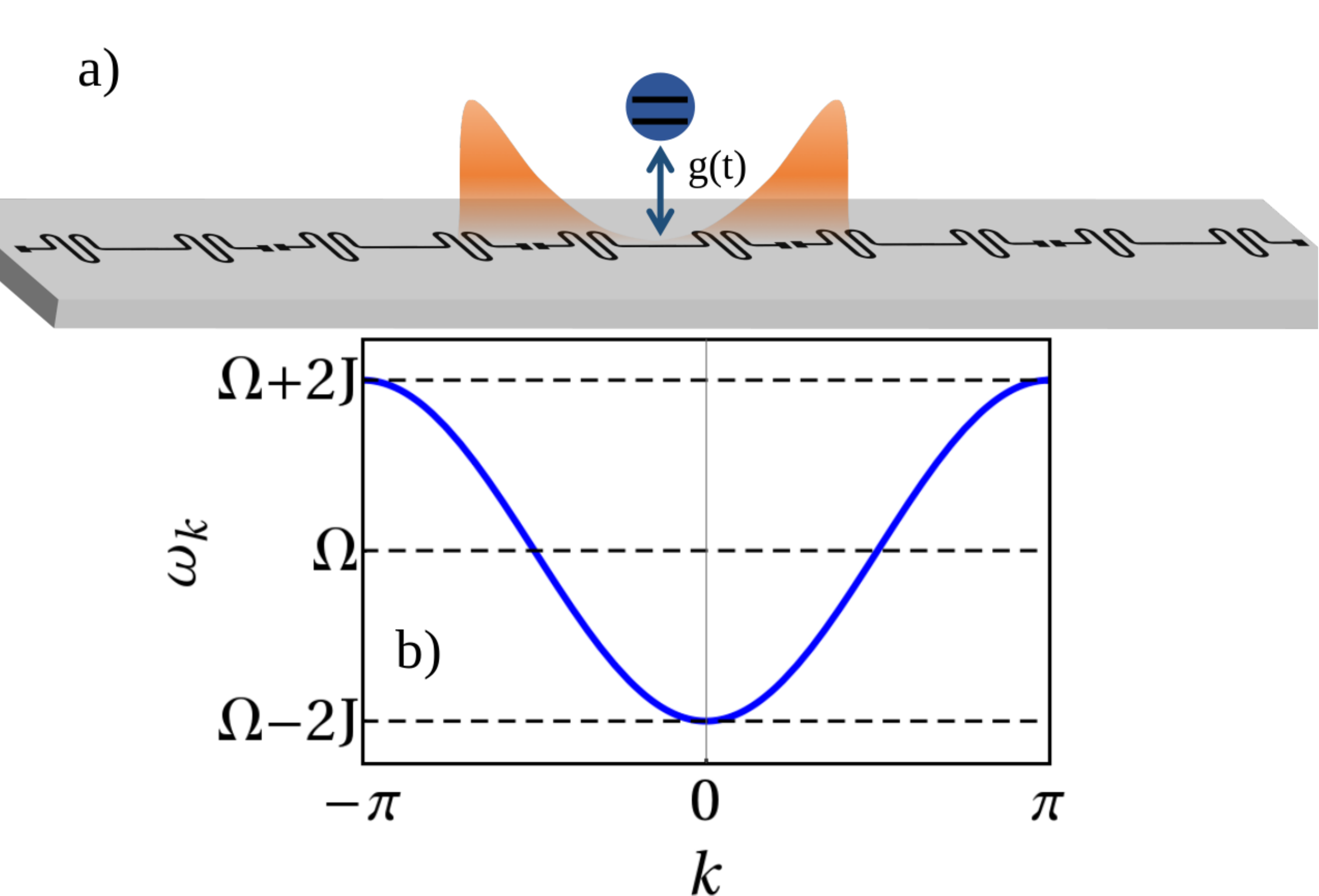}
\caption{(a) Sketch of the system. The 2LS-resonator interaction is $g(t)$. (b) Dispersion relation $\omega_k=\Omega-2J\cos k$ of the model given by Eq. \eqref{eq:Htbm}.}\label{Fig1}
\end{figure}
%%%%%%%%%%%%%%%%%%%%%%%%%%%%
The main novelty of our work is that it presents the first example of single-photon Fock states emitted from vacuum.
This is  different from the emission that occurs when a 2LS is ultrastrongly coupled to a cavity\ \cite{Niemczyk2010, Forn-Diaz2010}.
There, the emission is formed by photon pairs due to parity conservation\ \cite{DeLiberato2007,DeLiberato2009,Ciuti2005,Stassi2013,QiKai2018}.  In the geometry considered in this letter,  the existence of qubit-field bound states allows triggering single photons from vacuum.
Our theoretical proposal can be realized with superconducting circuits\ \cite{Gu2017}. Flux or transmon qubits ultrastrongly coupled to a superconducting waveguide\ \cite{Forn-Diaz2017,Puertas2018} should allow testing our results, enlarging the family of quantum field theory ideas\ \cite{Moore1970,Lhteenmki2013,Wilson2011} that superconducting circuits can emulate.
In particular the manifestation of  virtual photons, which is of current interest \cite{DeLiberato2017}.

%%%%%%%%%%%%%%%%%%%%%%%%%%%%%%%%%%%%%%%%%%
%%%%%%%%%%%%%%%%%%%%%%%%%%%%%%%%%%%%%%%%%%
%%%%%%%%%%%%%%%%%%%%%%%%%%%%%%%%%%%%%%%%%%
%%%%%%%%%%%%%%%%%%%%%%%%%%%%%%%%%%%%%%%%%%
\emph{Model.-}
We study the spin-boson model, a continuum of bosonic modes coupled to a 2LS\  \cite{Leggett1987}
\begin{equation}
  \label{Hsb}
  H = \Delta \sigma^+ \sigma^- + \sum_k \omega_k a_k^\dagger a_k
  +  \sigma_x \sum_k g_k (a_k^\dagger + a_k) \; ,
\end{equation}
The $\sigma^\pm$ are ladder operators of the 2LS and $\Delta$ is the excitation energy of the 2LS. The Pauli matrix $\sigma_x$ couples with strengths $g_k$ to the bosonic field operators  $\{a_k^\dagger, \,a_k\}$ in momentum space. We consider a dispersion relation $\omega_k= \Omega - 2 J \cos k$ (Fig. \ref{Fig1}b)) with $N$ momenta $k\in [-\pi,\pi)$ and a band edge that allows us to control the vacuum emission. This $\omega_k$ results from an one-dimensional array of cavities with nearest-neighbours coupling (Fig. \ref{Fig1}a))
\begin{equation}\label{eq:Htbm}
  H_{\rm array}=  \sum_{x=-N/2}^{N/2} (\Omega a_x^\dagger a_x - J( a_x^\dagger a_{x+1} + {\rm H.c.}) )\;,
\end{equation}
with bosonic operators in positions $\{a_x,\, a_x^\dagger\}$, resonator frequency $\Omega$ and hopping $J$. This choice of photonic band is not essential, but favours the numerical simulation.
The quantum emitter is coupled to a cavity at $x=0$, as in $H_{\rm coupling} = g \sigma_x (a_0 + a_0^\dagger)$, leading to $g_k=g/\sqrt{N}$ in Eq.\ \eqref{Hsb}.

If the coupling is sufficiently small, $g\ll \Delta $, the rotating-wave approximation (RWA) \cite{Cohen-Tannoudji1992} allows us to replace the interaction term with $H_{\rm coupling} \cong g(\sigma^+ a_0  + {\rm H.c.})$, which conserves the number of excitations.  
In this limit, the ground state has no excitations $|\text{GS}\rangle_{\rm RWA} = | 0; {\bf 0} \rangle$ and is the product of the 2LS ground state 
$|0\rangle$ ($|1\rangle$ is the excited state) and the zero-photon state of the waveguide $a_k| {\bf 0} \rangle =0.$ Therefore, under the RWA, the emitter is immune to the vacuum fluctuations of the bosonic field.
However,
 the RWA fails in computing the actual vacuum properties \cite{Loudon2006, Berman2006}.  
Beyond the RWA, the  ground state of \eqref{Hsb} contains excitations: $ \langle \text{GS} |a_x^\dagger a_x | \text{GS} \rangle \neq 0$, suggesting that the 2LS can convert fluctuations into radiated light. We investigate here the beyond-RWA vacuum emission of the spin-boson model \eqref{Hsb}.

% 

%%%%%%%%%%%%%%%%%%%%%%%%%%%%%%%%%%%%%%%%%%%%%
%%%%%%%%%%%%%%%%%%%%%%%%%%%%%%%%%%%%%%%%%%%%%
%%%%%%%%%%%%%%%%%%%%%%%%%%%%%%%%%%%%%%%%%%%%%
%%%%%%%%%%%%%%%%%%%%%%%%%%%%%%%%%%%%%%%%%%%%%
\emph{Theoretical tools.-}
The spin-boson model is not solvable, except for particular set of parameters and some limits, but matrix-product state (MPS) techniques can be used to obtain numerical results \cite{Peropadre2013b, Sanchez-Burillo2014, Sanchez-Burillo2015}, as explained in SM.
We contrast the numerical simulations with analytical approximations based on the polaron transformation \cite{Silbey1984, Bera2014, Camacho2016, Shi2018}.
This transformation is a disentangling operation $U_p$ that decouples the 2LS from the field
\begin{equation}
\label{Up}
  U_p =  \exp\left[ -\sigma_x \sum (f_k a_k^\dagger - f_k^* a_k) \right]\;.
\end{equation}
The parameters $f_k$ are obtained by minimizing the ground-state energy $E_{\text{GS}}$ within the polaron ansatz for the g.s. $|\text{GS}\rangle = U_p|0;\mathbf{0}\rangle$, giving the equations
\begin{equation}
\label{fk}
f_k =  \frac{g_k}{\Delta_r + \omega_k}, \;\mbox{and}\;
\Delta_r = \Delta e^{-2\sum_k |f_k|^2}.
\end{equation}
The simplified Hamiltonian $H_{p}=U_p^\dagger H U_p$ reads
\begin{align}
  \label{Hp}
  \nonumber
  H_{p} & = \Delta_r \sigma^+\sigma^- +  \sum_k 
  \omega_k a_k ^\dagger a_k - 2 \Delta_r \left( \sigma^+ \sum_k f_k a_k + {\rm H.c.} \right) \nonumber\\
  &-  2 \Delta_r \sigma_z \sum_{k,p} f_k^* f_p a_k^\dagger a_p \nonumber\\
  &+ \frac{\Delta}{2} + \sum_k (\omega_k |f_k|^2 -g_k^*f_k-f_k^*g_k) + {\rm h.o.t.}
\end{align}
Here, h.o.t. stands for higher-order terms ${\mathcal O} (f^3)$. The transformed Hamiltonian conserves the number of excitations and can be treated analytically\ \cite{Shi2018}.

The renormalization of the 2LS energy $\Delta_r$ is a consequence of the coupling of a discrete quantum system to a continuum \cite{Leggett1987}  (see SM).
   According to the polaron picture, most correlations are captured by the unitary transformation of a product state $|\text{GS}\rangle_{\rm p}= |0 ;{\bf 0}\rangle = U_{\rm p}^\dagger |\text{GS} \rangle $. Then, $U_{\rm p}$ plays a similar role to the Bogoliubov transformations\ \cite{Altland2010} used for finding the normal modes which account for the radiation in the Hawking, Unruh or Casimir effects \cite{Nation2012}.

%%%%%%%%%%%%%%%%%%%%%%%%%%%%%%%%%%%%%%%%%%%%%
%%%%%%%%%%%%%%%%%%%%%%%%%%%%%%%%%%%%%%%%%%%%%
%%%%%%%%%%%%%%%%%%%%%%%%%%%%%%%%%%%%%%%%%%%%%
%%%%%%%%%%%%%%%%%%%%%%%%%%%%%%%%%%%%%%%%%%%%%
\emph {Spectrum of the spin-boson model.- }
\begin{figure*}[tbh!]
\includegraphics[scale=0.25]{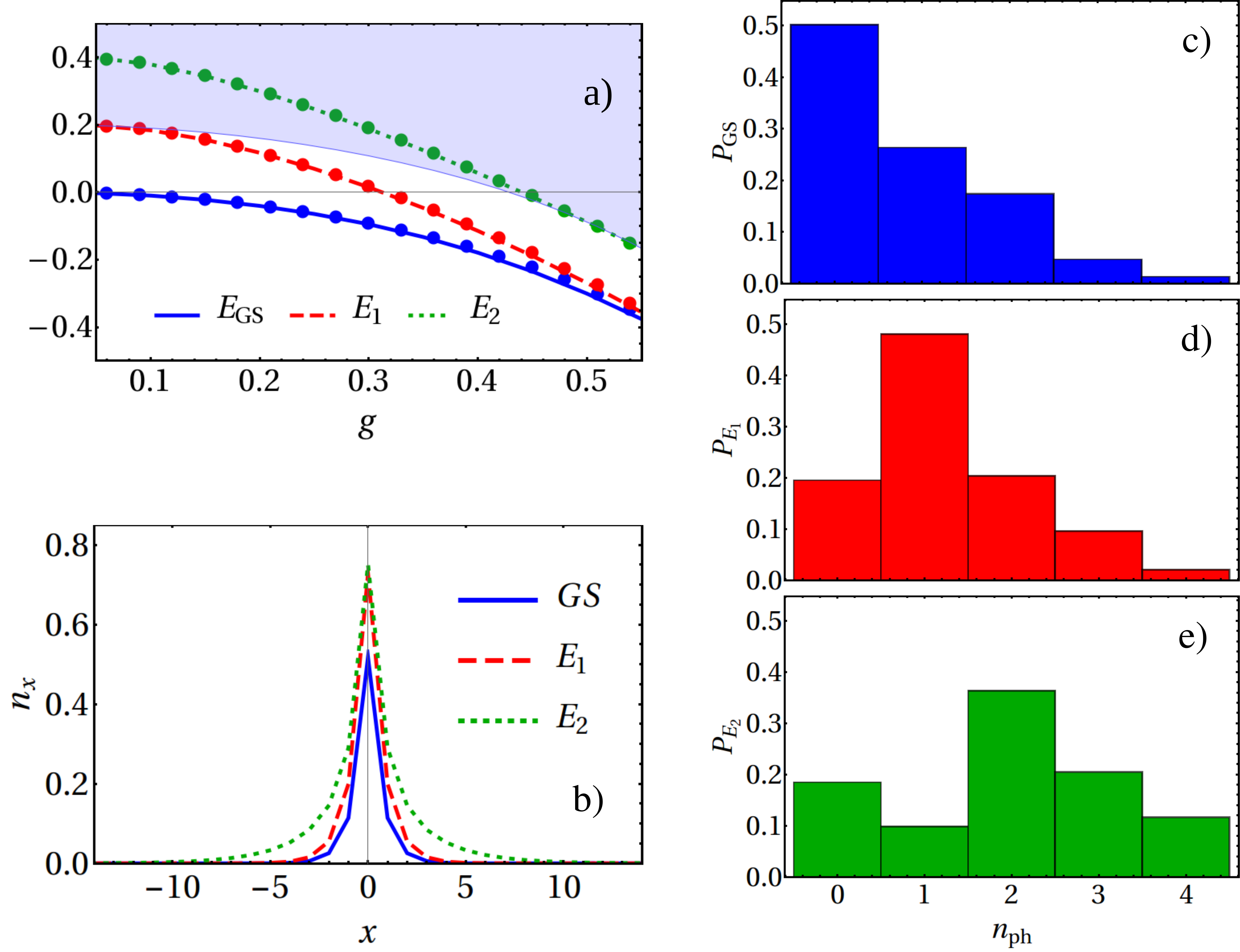}
\caption{Bound states. (a) Eigenenergies as a function of $g$ for $\Delta=0.3$. Continuous lines stand for the MPS simulations and the points for the polaron ansatz. (b) Bbound states in position space for $g=0.5$ and $\Delta=0.3$. (c), (d), and (e) Histograms with the weights in the $n_\text{ph}$-photon sector for $\ket{\text{GS}}$, $\ket{E_1}$, and $\ket{E_2}$. Same parameters as in panel (b). The parameters defining the photonic waveguide are  $\Omega=1.0$ and $J=0.4$. The lattice length is $N=400$.}\label{Fig2}
\end{figure*}
The spectrum of the Hamiltonian \eqref{Hsb} is essential to understand the dynamics of vacuum-induced photon emission.
The photonic band edge causes the appearance of photon bound states: localized excitations around the 2LS \cite{Sanchez-Burillo2014,Shi2009,Lombardo2014,Shi2016,Calajo2016,Sanchez-Burillo2017}. We classify those states according to their parity $\Pi = \exp \big ( i\pi(\sigma^+\sigma^- + \sum_k a_k^\dagger a_k) \big )$, which is a conserved quantity\ \eqref{Hsb}, $[\Pi,H]=[\Pi,H_p]=0.$ More precisely, the ground state $\ket{\text{GS}}$ and the second bound state $\ket{E_2}$ are the first and second eigenstates with even parity $\Pi=+1.$ The first bound state $\ket{\Psi_1}$ is the lowest eigenstate with odd parity $\Pi=-1$. $\ket{E_1}$ and $\ket{E_2}$ have a well-defined number of particles in the RWA limit (1 and 2 respectively).% and sits below the band with one propagating photon $E_1\leq \epsilon_k(GS)\equiv E_{GS} + \omega_k$, just as in the RWA model\ \cite{Longo2010,Longo2011,Shi2016,Calajo2016,Sanchez-Burillo2017}.

We compute these states using both MPS and the polaron Hamiltonian. Parity can be imposed during the MPS minimization of $H$; in the second case, we project the polaron Hamiltonian (Eq.\ \eqref{Hp}) onto spaces with fixed number of excitations, where it is numerically diagonalized. Fig. \ref{Fig2}a) shows the energy of the ground state  $E_{\text{GS}}$ and of the first two bound states, $E_1$ and $E_2$, as a function of the coupling $g$. Note the excellent agreement between MPS (solid line) and the polaron Hamiltonian calculations (dots). Note also how the first bound state lays just below the one-photon band (gray band) $E_1\leq \epsilon_k(\text{GS})\equiv E_{\text{GS}} + \omega_k$, just as in the RWA model\ \cite{Longo2010,Longo2011,Shi2016,Calajo2016,Sanchez-Burillo2017}. The second excited bound state $E_2$ enters the band of propagating single photons. There may be other bound states, but the overlap with propagating photon bands of similar parity turn these bound states (which within the RWA would be perfectly localized) into resonances with a finite lifetime\ \cite{Sanchez-Burillo2014}. Further comparisons between results obtained using MPS and the polaron transformation are given in SM.

We have also analyzed the bound state MPS wavefunctions, $\ket{E_1}$ and $\ket{E_2}$. These states are localized around the 2LS, as seen in Fig. \ref{Fig2}b), which renders the number of photons in real space $\langle n_x \rangle = \langle a_x^\dagger a_x \rangle$. Interestingly, since the MPS produces wavefunctions in the original frame of reference---i.e. after applying $U_p$ onto the polaron states---, we find that these states are actual superpositions of different numbers of photons, as seen in Fig. \ref{Fig2}c-d-e). The overall superposition preserves the parity of the state but, say, a bound state with two excitations can have a nonzero overlap with a single-photon component.
%

%%%%%%%%%%%%%%%%%%%%%%%%%%%%%%%%%%%%%%%%%%%%%
%%%%%%%%%%%%%%%%%%%%%%%%%%%%%%%%%%%%%%%%%%%%%
%%%%%%%%%%%%%%%%%%%%%%%%%%%%%%%%%%%%%%%%%%%%%
%%%%%%%%%%%%%%%%%%%%%%%%%%%%%%%%%%%%%%%%%%%%%
\begin{figure}[tbh!]
\includegraphics[scale=0.26]{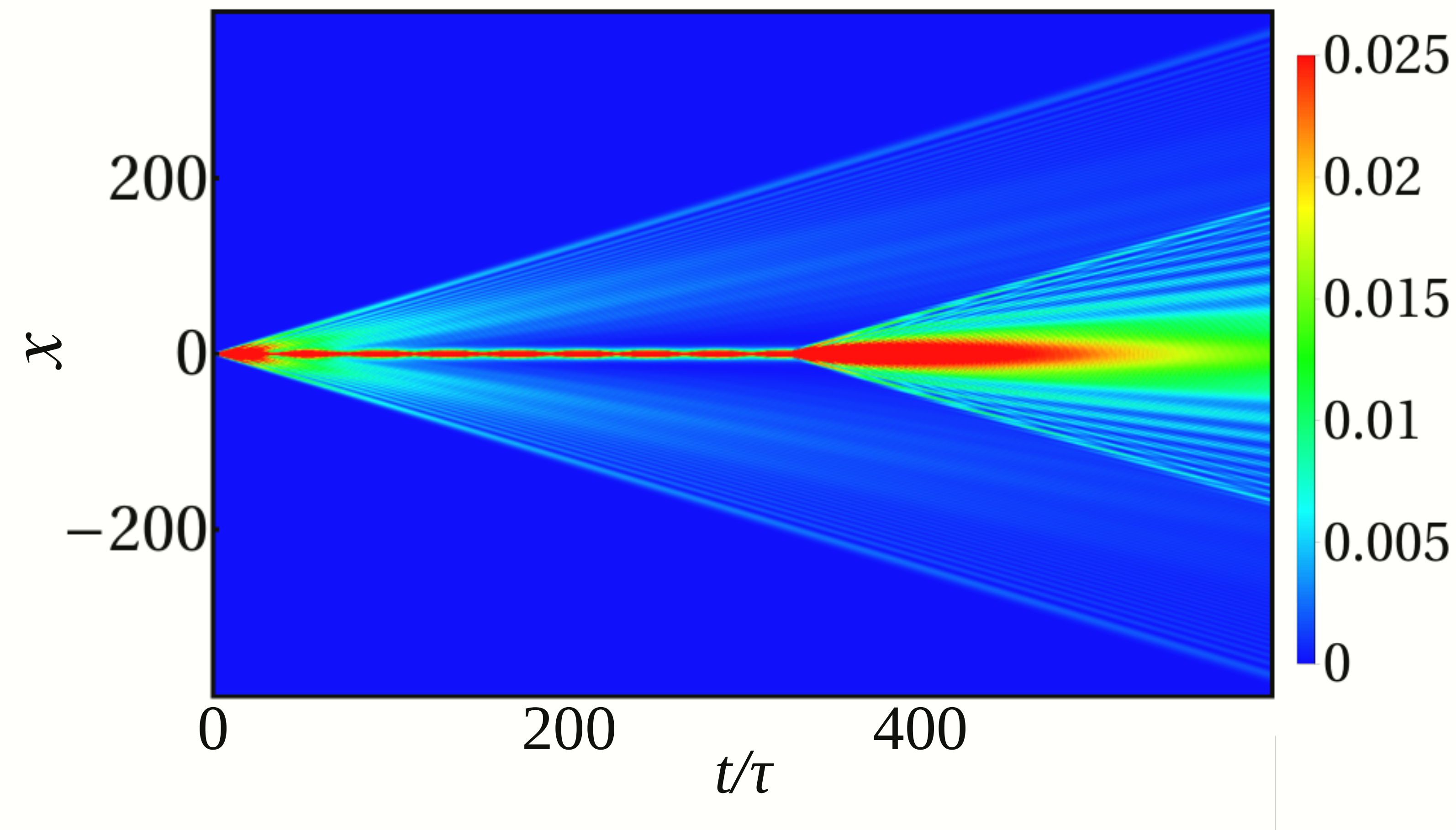}
\caption{Number of photons as a function of time and position for the quenching protocol: the initial state is the trivial vacuum $\ket{\Psi(t=0)}=\ket{0;0}$ and the coupling is switched on at $t=0$. We switch $g$ off at $t_\text{off}/\tau=350$. The system emits a wavepacket at $t=0$. At $t=t_\text{off}$ it radiates again. $g=0.5$ after the initial quench and $\Delta=0.3$. The rest of parameters are as in Fig. \ref{Fig2}}\label{Fig3}
\end{figure}
%It is time to show the light emission by probing the vacuum with a 2LS.

\emph {Emission by quenching the vacuum .-}
To convert vacuum excitations into emitted light, we consider a nonadiabatic protocol where the light-matter coupling strength is rapidly switched on and off.
 An alternative protocol, probably more amenable to experimental study, is to abruptly modify the qubit excitation energy $\Delta$ from a value that is strongly detuned from the photonic band $g/(\Delta(t<0)-\omega_k) \ll 1$ to a value  $\Delta(t>0) \sim \omega_k$ while keeping a constant coupling $g$.
Both methods are theoretically equivalent, since both ground states are the same up to an error ${\rm exp} [
  -g^2/\Delta]$ that can be made arbitrarily small SM.
  In what follows, we analyze the coupling quench, which is simpler to describe both analitically and numerically, since the decoupled limit corresponds to $g=0$, while in the other case full decoupling occurs for infinite $\Delta$.
We begin with an unexcited 2LS with $g(t<0)=0$ and switch on the coupling strength to a value $g(t=0)>0$ beyond the RWA regime.
The 2LS immediately begins to emit light to accommodate its new ground state. The emitted photons form a wavepacket that travels with speed ${\rm max}_k(\partial_k \omega_k)$. After some time the 2LS is no longer emitting and the wavepacket leaves a cloud around the 2LS. We then suddenly switch off the coupling at $t=t_\text{off}$ and a second vacuum emission takes place.

We simulate the dynamics described in the previous paragraph with MPS. The initial state is the trivial vacuum $\ket{ \Psi (t=0) } = \ket{ 0 ; {\bf  0} }$, which corresponds to the uncoupled case $g=0$, and $\ket{\Psi(t)}$ evolves under \eqref{Hsb} with $g$ within the ultrastrong. In Fig. \ref{Fig3}, we plot the photon number $n_x = \langle a_x^\dagger a_x \rangle $ along waveguide, as a function of time $t$ and position $x$. Note how all perturbations emerge from the 2LS position. We switch off the coupling once the travelling photons are far from the emitter. We choose $t_\text{off}=350 \tau$, being $\tau$ the spontaneous decay rate of the 2LS given by the Fermi's golden rule: $\tau\equiv J\sin (k_0) /g^2$, with $k_0$ such that $\omega_{k_0}=\Delta$. At this point $g(t_\text{off})=0$ and we witness the second photon emission event.
Notice that photons propagate with different velocities because of the nonlinearity of the dispersion relation $\omega_k$ (see below  Eq. \eqref{Hsb}).

The whole process admits a simple description in the polaron picture. The state before the quench is
\begin{equation}
\label{psi0}
\ket{\Psi (t=0)}_p = U_p^\dagger\ket{0 ; {\bf 0}} = \frac{1}{\sqrt{2}}  \big ( \ket{0 ; {\bf \alpha_+}} + \ket{1; {\bf \alpha_-}} \big )\,.
\end{equation}
This is a superposition of even and odd cat states $|\alpha_\pm \rangle \equiv e^{ \sum (f_k a_k^\dagger - f_k^* a_k)}\ket{\bf 0} \pm e^{ -\sum (f_k a_k^\dagger - f_k^* a_k)} \ket{\bf{0}}$.  In the limit of weak amplitudes, $|\alpha_\pm \rangle $ tend to one- and two-photon states respectively,\ \cite{Gheeraert2017} and the wavefunction can be written using bound and propagating states. Asymptotically in time, the state has the form
\begin{align}
\label{psipt}
\ket{\Psi (t)}_p
& = 
c_{0,0}(t) \ket{0;{\bf 0}} + c_{0,2}(t) \ket{E_2}
\\ \nonumber
&+
c_{1,1}(t) A_{1\gamma}^\dagger\ket{E_1} + c_{2,0}(t)
A_{2\gamma}^\dagger\ket{0;{\bf 0}}+ \ldots
\end{align}
This wavefunction allows four possible outcomes: the system goes to (i) the ground state or (ii) to $\ket{E_2}$ with no emission; (iii) it relaxes to the first odd bound state $\ket{E_1}$ emitting a wavepacket $A_{1\gamma}^\dagger$ with one photon, or (iv) it relaxes to the ground state emitting two photons $A_{2\gamma}^\dagger.$ Note that when we write this wavefunction in the laboratory basis $\ket{\Psi(t)} = U_p\ket{\Psi(t)}_p,$ the structure of the state is preserved, because the polaron transformation is local in space $[A_{1,2\gamma}^\dagger,U_p]=0$  SM.

We have tested numerically that Eq.\ \eqref{psipt} captures the vacuum-triggered emission.
%
%First, $\ket{\Psi(t)}_p$ explains the 2LS oscillations $P_{\rm qb}(t) = \braket{\Psi (t) | \sigma^+ \sigma^- | \Psi (t)}$,  shown in Fig. \ref{Fig3}b). These oscillations show the interference of the amplitudes $|\text{GS}\rangle$ and $|E_2\rangle$. They estabilize after a transient  time of the order of the qubit relaxation time, $t\geq \tau$, and persist until the coupling is switched off at $t_\text{off}$, when $P_{qb}$ gets frozen.

The simulations confirm that the system emits photons mainly in two channels: (i) one photon on the first excited odd bound state and (ii) two photons on the ground state, as predicted by Eq. \eqref{psipt}.  This is shown in Fig. \ref{Fig4}a), where we plot the number of photons $n_x$ at time $t/\tau=250$ and the single-photon $n_x^{(1)}= | \langle \Psi (t)| a_x^\dagger | E_1 \rangle|^2$ and two-photon contributions $n_x^{(2)}= 2\sum_{x^\prime} | \langle \Psi (t)| a_x^\dagger a_{x^\prime}^\dagger | \text{GS} \rangle|^2$. As seen, $n_x$ is well approximated by the sum of both wavepackets $n_x^{(1)}+n_x^{(2)}$.

$\ket{\Psi(t)}_p$ also explains the second photon emission event. In this case, once we switch off the couplings, the bound states become unstable and decay, releasing their photonic components in the form of propagating photons.
  These come from the three first terms in Eq. \eqref{psipt}.
  Two main features stand out.
  First, more power is radiated than in the first quench.
  This is because, in this second quench, excited bound states also radiate.
  Second, the radiated flying photons are \emph{slower}.
This is 
because  the bound states are spectrally close to the photonic-band minimum, so radiation occurs mainly into slow photons.
 The distribution of this emitted light for each bound state matches the statistics in Figs. \ref{Fig2}c-e). 

The simulations prove that $\ket{\Psi(t)}_p$ also explains the 2LS dynamics SM.

We can control the vacuum induced emission, for instance selecting the one-photon channel, by playing with the relative values of the band gap $\omega_{k=0}$ and the bound state energy $E_1-E_{\text{GS}}$. The energies of the radiating states with one and two flying photons are $\epsilon_k(E_1)=E_{1}(g)+\omega_{k}$ and $\epsilon_{k_1,k_2}(\text{GS})=E_{\text{GS}}+\omega_{k_1}+\omega_{k_2}$, with respective minima $E_1(g)+\omega_{k=0}$ and $E_{\text{GS}}+2\omega_{k=0}$. If we place the emitter in the band gap $\omega_{k=0}\gg \Delta$, the energies $\epsilon_k(E_1)$ become closer to the 2LS resonance with respect to $\epsilon_{k_1,k_2}(\text{GS})$, so the two-photon component is strongly suppressed ($|c_{2,0}|\simeq 0$ in Eq. \eqref{psipt}). The selectivity of this process is confirmed in Fig. \ref{Fig4}b), where the considered band gap is five times larger than in Fig. \ref{Fig4}a) and all other parameters are equal. The final state has a negligible overlap with $\ket{E_2}$ and the distribution of photons $P_{E_n}$ contains less than 1\% of components with $n_\text{ph} \ge 2$. The state before the second quench is faithfully reconstructed by just its single-photon component $A_{1\gamma}^\dagger\ket{E_1},$ and as a result, the second emission is also well approximated by one photon.

\begin{figure}
\includegraphics[scale=0.17]{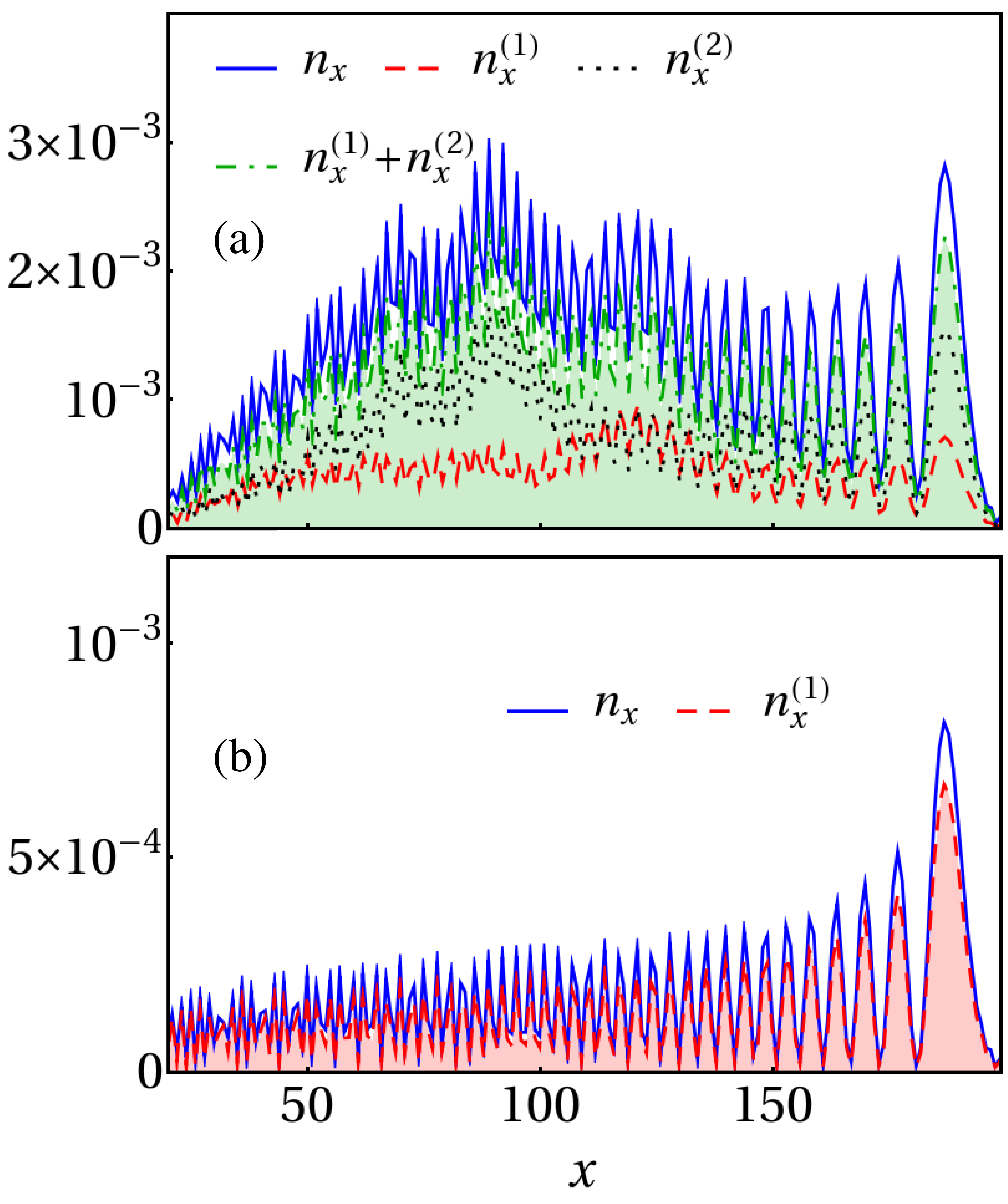}
\caption{(a) Number of photons at time $t/\tau=250$ after the quantum quench described in Fig. \ref{Fig3}. We can approximate the field with one- and two-photon components. All the parameters are those of the previous figures (see Figs. \ref{Fig2} and \ref{Fig3}). (b) Same as before, increasing the energy of the resonators $\Omega$ such that the band gap is now five times larger: $\Omega=1.8$ (remind that the band gap is $\Omega-2J$). The system emits a single-photon packet.}\label{Fig4}
\end{figure}

%%%%%%%%%%%%%%%%%%%%%%%%%%%%%%%%%%%%%%%%%%%%%
%%%%%%%%%%%%%%%%%%%%%%%%%%%%%%%%%%%%%%%%%%%%%
%%%%%%%%%%%%%%%%%%%%%%%%%%%%%%%%%%%%%%%%%%%%%
%%%%%%%%%%%%%%%%%%%%%%%%%%%%%%%%%%%%%%%%%%%%%
\emph { Conclusions.-}
In this work we have studied the dynamics of vacuum fuctuations in \emph{ultrastrong} waveguide-QED setups. More precisely, we have shown that the nonlinearity of a 2LS, combined with a nonperturbative coupling to a bosonic field, can be used to create a vacuum-triggered single-photon emitter. In other words, we discuss the ultimate limit of quantum nonlinear optics as driven by vaccum fluctuations \cite{Chang2014}. %Besides, as the emitted state is nonGaussian, it can potentially have applications in several quantum information protocols \cite{Gottesman2001,Cerf2005,Ghose2005,Weedbrook2012,Hughes2016}.
Our proposal is analogous in spirit to other quantum-field theory inspired proposals, such as the dynamical Casimir effect, which work with nonperturbative and nonadiabatic changes of the theory. In contrast to those experiments, we have shown a minimum setup which extracts  single photons from vacuum, using bound states as mediators of these processes. It is important to remark that this whole study can be repeated using a resonator instead of a 2LS. In this case, all of the features above disappear, as the emission has a Gaussian statistics that 
 are not Fock states \cite{Genoni2010}.

Our proposal and the conditions in this work can be realized in current circuit-QED devices with superconducting qubits that are ultrastrongly coupled to open transmission lines \cite{Forn-Diaz2017,Puertas2018}. In this exciting platform, state-of-the-art measurement techniques would allow for a detailed reconstruction of the photon wavepackets \cite{Menzel2010,Eichler2011}.

%%%%%%%%%%%%%%%%%%%%%%%%%%%%%%%%%%%%%%%%%%%%%
%%%%%%%%%%%%%%%%%%%%%%%%%%%%%%%%%%%%%%%%%%%%%
%%%%%%%%%%%%%%%%%%%%%%%%%%%%%%%%%%%%%%%%%%%%%
%%%%%%%%%%%%%%%%%%%%%%%%%%%%%%%%%%%%%%%%%%%%%
\emph { Acknowledgments.-}
%%%%%%%%%% acknowledgments
We acknowledge the Spanish Ministerio de Ciencia, Innovación y Universidades within project MAT2017-88358-C3-1-R and FIS2015-70856-P and the Aragón Government project Q-MAD and CAM PRICYT Research Network QUITEMAD+ S2013/ICE-2801. EU-QUANTERA project SUMO is also acknowledged. Eduardo Sánchez-Burillo acknowledges ERC Advanced Grant QENOCOBA under the EU Horizon 2020 program (grant agreement 742102).

\newpage

\begin{widetext}
	\widetext
	\onecolumngrid
	% \vspace*{\columnsep}
	% % \vspace*{\columnsep}
	\begin{center}
		\textbf{\large Supplemental Material: Chiral quantum optics in photonic sawtooth lattices}
	\end{center}
	\vspace{\columnsep}
	\vspace{\columnsep}
	
	\twocolumngrid
\end{widetext}
%%%%%%%%%% Merge with supplemental materials %%%%%%%%%%
%%%%%%%%%% Prefix a "S" to all equations, figures, tables and reset the counter %%%%%%%%%%
\setcounter{equation}{0}
\setcounter{figure}{0}
\setcounter{section}{0}
\makeatletter

%SM for figures
\renewcommand{\thefigure}{SM\arabic{figure}}
\renewcommand{\thesection}{SM\arabic{section}}  
\renewcommand{\theequation}{SM\arabic{equation}}  

\title{Single photons by quenching the vacuum
\\
Supplemental Material}

This supplemental material is structured in sections: (i) we first prove that both quenching protocols (coupling and detuning) are equivalent, (ii) we briefly summarize the basics of matrix-product states, (iii) we test the polaron transformation, and (iv) we prove that the polaron ansatz explains the qubit dynamics.

%\pacs{42.25.Bs, 41.20.Jb, 42.79.Ag, 78.66.Bz}

\maketitle
%%%%%%%%%%%%%%%%%%%%%%%%%%%%%%%%%%%%%%%%%%%%%
%%%%%%%%%%%%%%%%%%%%%%%%%%%%%%%%%%%%%%%%%%%%%
%%%%%%%%%%%%%%%%%%%%%%%%%%%%%%%%%%%%%%%%%%%%%
%%%%%%%%%%%%%%%%%%%%%%%%%%%%%%%%%%%%%%%%%%%%%

%%%%%%%%%%%%%%%%%%%%%%%%%%%%%%%%%%%%%%%%%%%%%
%%%%%%%%%%%%%%%%%%%%%%%%%%%%%%%%%%%%%%%%%%%%%
%%%%%%%%%%%%%%%%%%%%%%%%%%%%%%%%%%%%%%%%%%%%%
%%%%%%%%%%%%%%%%%%%%%%%%%%%%%%%%%%%%%%%%%%%%%

\section{Equivalence of the two protocols}
\label{app:gap}

\begin{figure}[h]
\includegraphics[scale=0.3]{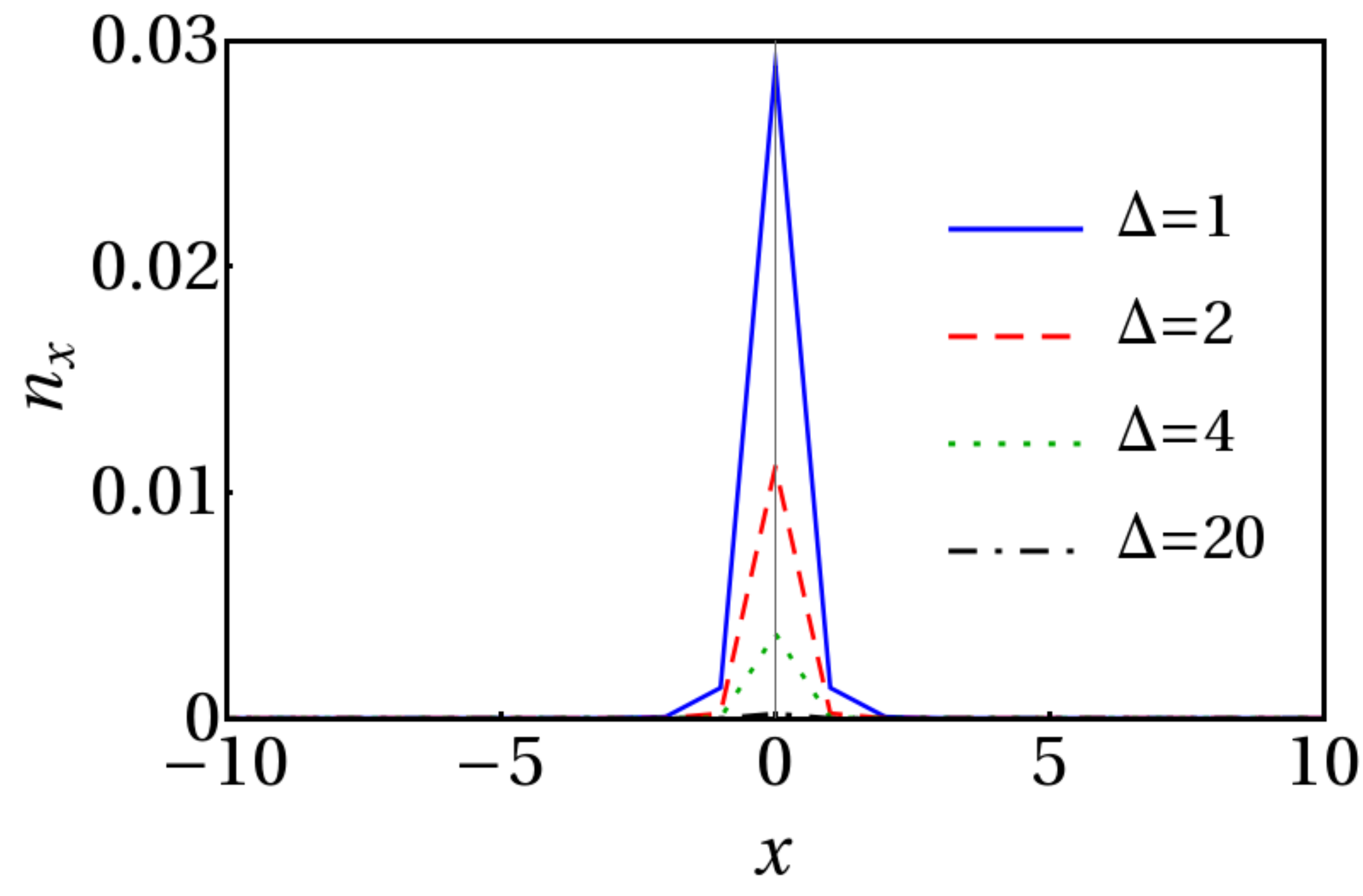}
\caption{Photon population in the ground state: $\langle \text{GS} | a_x^\dagger a_x | \text{GS} \rangle$.  We use $g=0.3$, $\Omega=1$, $J=0.4$. The lattice length is $N=400$.  Different values for $\Delta$ are shown in the legend.}
\label{fig:nxgs}
\end{figure}

We show here that both protocols, namely the quenching in the coupling and the detuning-tuning  of the 2LS frequency, yield  an equivalent dynamics.
The first part of  our demonstration compares the initial ground states in both protocols. 
In the quenching one, since $g(t<0)=0$,  this is the trivial product of the 2LS ground state and the zero-photon state  $|0; {\bf 0} \rangle$.  
On the other hand, in the detuning protocol, the ground state can be approximated with the help of the Polaron transformation as $| \text{GS} \rangle = U_p |  |0; {\bf 0} \rangle$.  In the large -detuned limit, \emph{i.e.} whenever $\Delta \gg \omega_k$, $\forall k$, the varational parameters tend to  $f_k \cong g_k / \Delta$.  Therefore, the fidelity between both  ground states is:
\begin{equation}
{\mathcal F}= | \langle  |0; {\bf 0}  | U_p |0; {\bf 0} \rangle|^2 = e^{- \sum_k |f_k|^2 } \to e^{
 -(g/\Delta)^2} \;,
\end{equation}
which  can be arbitrarily close to one.
Within the range studied in this paper, $g=0.3-0.5$ with a detuned qubit of $\Delta = 10 \Omega$, which can be of order of 10 GHz  in a superconducting architecture yields $\mathcal F \cong  0.997$. Further characterization is given by the photon number in the g.s. as a function of the detuning.  In figure \ref{fig:nxgs} , we see how the photon population downs to zero as the detuning increases.

\begin{figure}[h]
\includegraphics[scale=0.26]{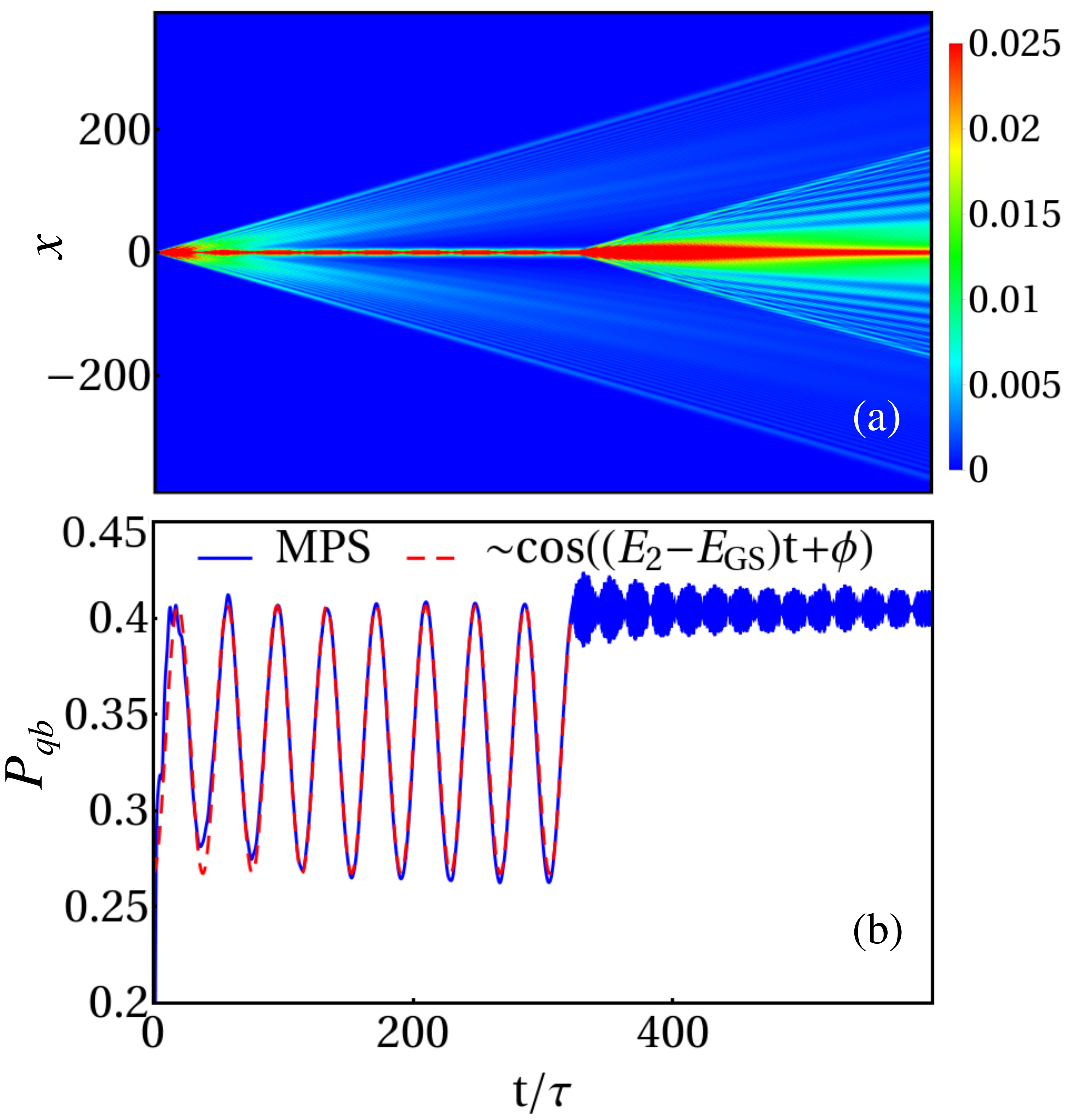}
\caption{Dynamics analogous to Fig. 3 of the main text for the \emph{detuning} protocol (see discussion in Sect. \ref{app:gap} of this SM). (a) Number of photons as a function of both time and position. At $t_\text{off}/\tau=350$, we set $\Delta_f=10$, well outside the photonic band and the bound states radiate photons. (b) Population of the excited state of the 2LS. After the quench, the population still evolves (compare to Fig. 3 of the main text). Apart from the detuning after the final quench $\Delta_f$, the parameters are those of Fig. 3 of the manuscript.}
\label{fig:dynamics_gap}
\end{figure}

This shows the equivalence in the first quench of the protocol.  To finish our demonstration we show the full dynamics, also after the second quench 
for the he emitted field and the dynamics of the 2LS population computed with MPS in in Fig. \ref{fig:dynamics_gap}. The emitted field behaves qualitatively as in the coupling-decoupling protocol considered in the main text (see Fig. 3a) of the main text). The 2LS population here, however, still evolves, after the quench, depicting some oscillations. The amplitude of these oscillations decreases as $\Delta_f$ increases (not shown) which does not modify  our conclusions.

\section{Matrix-product states}
\label{app:mps}

As we indicated in the main text, we use the MPS technique to compute the eigenstates and dynamics of the system. Let us justify why we can do it.

Our initial condition is the state with no excitations. After the nonadiabatic driving, the state is a combination of some of the lowest-energy states with a few flying photons (see Eq. (7) of the main text). As we are in the low-energy sector, we expect our state to fulfill the area law \cite{Eisert2010}, that is, it will be slightly entangled. Therefore, we may use matrix-product states \cite{Vidal2003,Vidal2004,Verstraete2004,Ripoll2006,Verstraete2008}, since it is valid for 1D systems when the entanglement is small enough. This ansatz has the form
\begin{equation}
\ket{\Psi} = \sum_{s_i \in \{1,d_i\}} \mathrm{tr}\left[
\prod A_i^{s_i}\right] \ket{s_1,s_2,\ldots,s_L}.
\label{eq:mps}
\end{equation}
This state is constructed from $L$ sets of complex matrices $A_i^{s_i} \in M[\mathbb{C}^{D}]$, where each set is labeled by the quantum state $s_i$ of the corresponding site. The local Hilbert space dimension $d_i$ is infinity, since we are dealing with bosonic sites. However, during the dynamics, processes that create multiple photons are still highly off-resonance. Then, we can truncate the bosonic space and consider states with $0$ to
$n_{max}$ photons per cavity. So, the composite Hilbert space is $\mathcal{H}=\bigotimes_i \mathbb{C}^{d_i}$, where the dimension is $d_i=n_{max}+1$ for the empty resonators and $d_{i_0}=2(n_{max}+1)$ for the cavity with the 2LS. We thus expect the state of the photon-2LS system to consist of a superposition with a small number of photons. In our simulations, we checked that $n_\text{max}=5$ is enough for good convergence.

The number of variational parameters is $(L-1)D^2(n_{max}+1) + 2D^2(n_{max}+1)$. In general, the matrix size $D$ increases exponentially with $L$ for typical states, whereas its dependence is polynomial if the entanglement is small enough, which usually occurs for low-energy states. Thus, the number of parameters increases polynomially with $L$ for slightly entangled states. In our simulations, $D\simeq 10-20$ proved to be enough.

Our work with MPS relies on three different algorithms. (i) The most basic one is to create the trivial initial state $\ket{\Psi(t=0)}=\ket{0;\mathbf{0}}$, which is actually a product state. This kind of state can be reproduced using matrices of bond dimension $D=1$, so each matrix is just a coefficient $A_i^{s_i}=\delta_{s_i1}$. (ii) The second algorithm is to compute expectation values from MPS. This amounts to a contraction of tensors that can be performed efficiently \cite{Ripoll2006}, and allows us to compute single-site operators $\langle a^\dagger_i a_i\rangle$, $\langle \sigma_z\rangle$, for instance. (iii) Finally, we can also approximate time evolution, both in real and imaginary times, repeatedly contracting the state with an approximation of the unitary operator $\exp(-iH\Delta t)$ for short times, and truncating it to an ansatz with a fixed $D$. Since our problem does just contain nearest-neighbour interactions, it is sufficient to rely on a third-order Suzuki-Trotter formula \cite{Suzuki1991}.
Taking imaginary times, we can obtain the ground state and excited states by solving the equation $i\tfrac{d}{dt}P\ket{\Psi}=PHP\ket{\Psi}$. Here, $P$ is either the identity (for the ground state) or a projector that either selects a well defined quantum number (\emph{e.g.} the parity) or projects out already computed states (for instance the ground and first-excited states). In either case, given a suitable initial state, the algorithm converges to the lowest-energy state of the Hamiltonian in the subspace selected by $P$.

\section{Further polaron tests}
\label{app:polaron}

We complement the main text with more calculations within the polaron picture.

\begin{figure}[tbh!]
\includegraphics[scale=0.26]{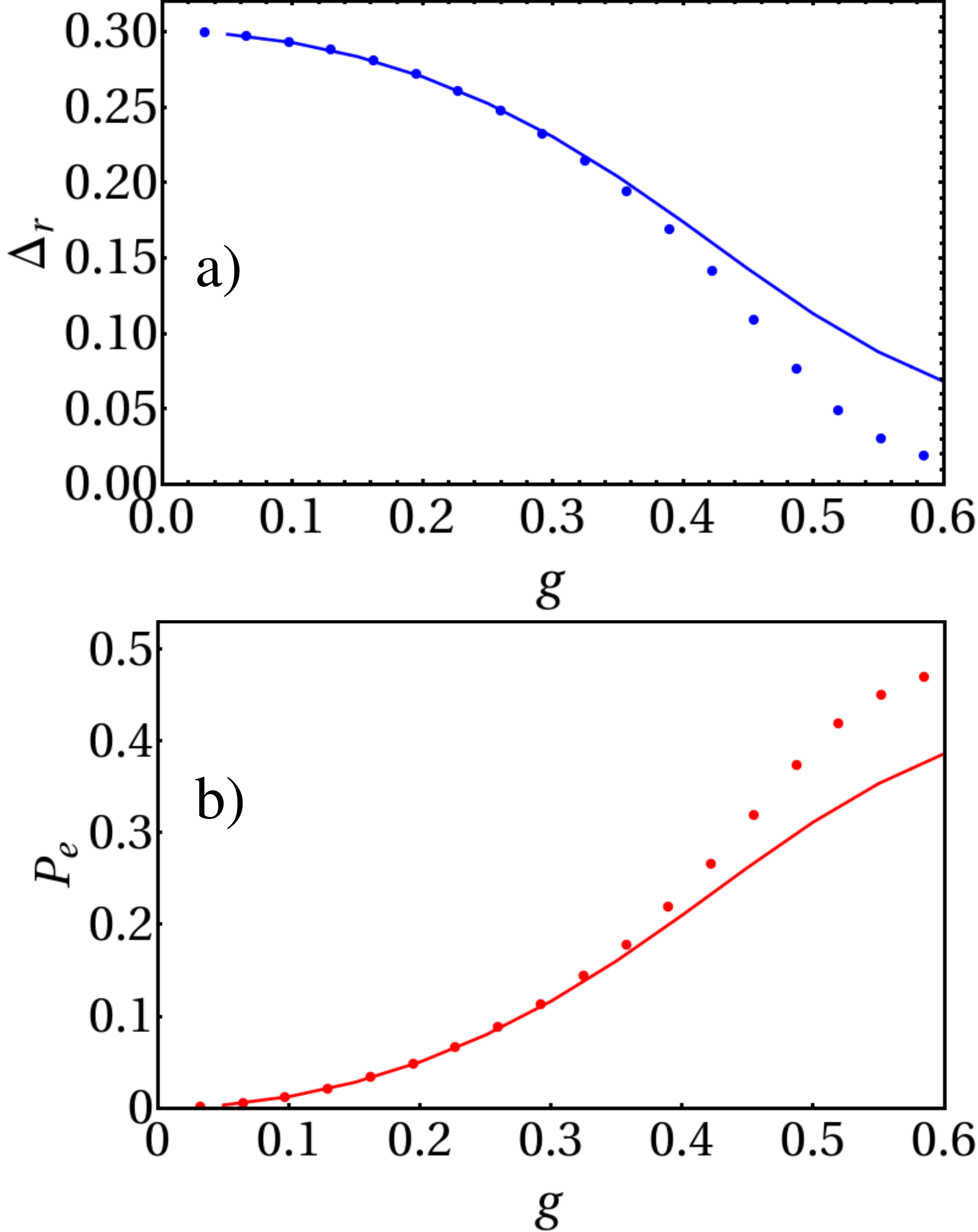}
\caption{(a) The renormalized frequency $\Delta_r$ as given by Eq. (4) in main text. (b) Populaton of the excited state of the 2LS in the ground state. In both plots, the continuous lines render results from the MPS simulations, whereas the dots represent the results obtained with the polaron ansatz. The parameters $\Omega$, $J$, and $\Delta$ are those of Fig. 3 of the manuscript.}\label{fig:Dr}
\end{figure}

First, we show the $g$-dependence of $\Delta_r$, Eq. (4). Notice that it is a self consistent equation, so it is solved numerically.  The results are given in Fig. \ref{fig:Dr}a).  We also show the probability for the 2LS to be excited, which can be directly computed from the frequency renormalization, namely:
\begin{equation}
\label{Pe}
P_e = \frac{\langle gs | \sigma_z | gs \rangle + 1}{2}= \frac{1-\Delta_r/\Delta}{2} \, .
\end{equation}
The results are plotted in Fig. \ref{fig:Dr}b). We see the reduction of $\Delta_r$, which resembles that in the spin-boson model. The reason, as seen in Eq. \eqref{Pe}, is that the 2LS and the field are hybridized in the ultrastrong regime, so $P_{qb} > 0$.  On top of that, we check that for $g \leq 0.5$ both the polaron transformation and MPS results agree.

\begin{figure}[tbh!]
\includegraphics[scale=0.38]{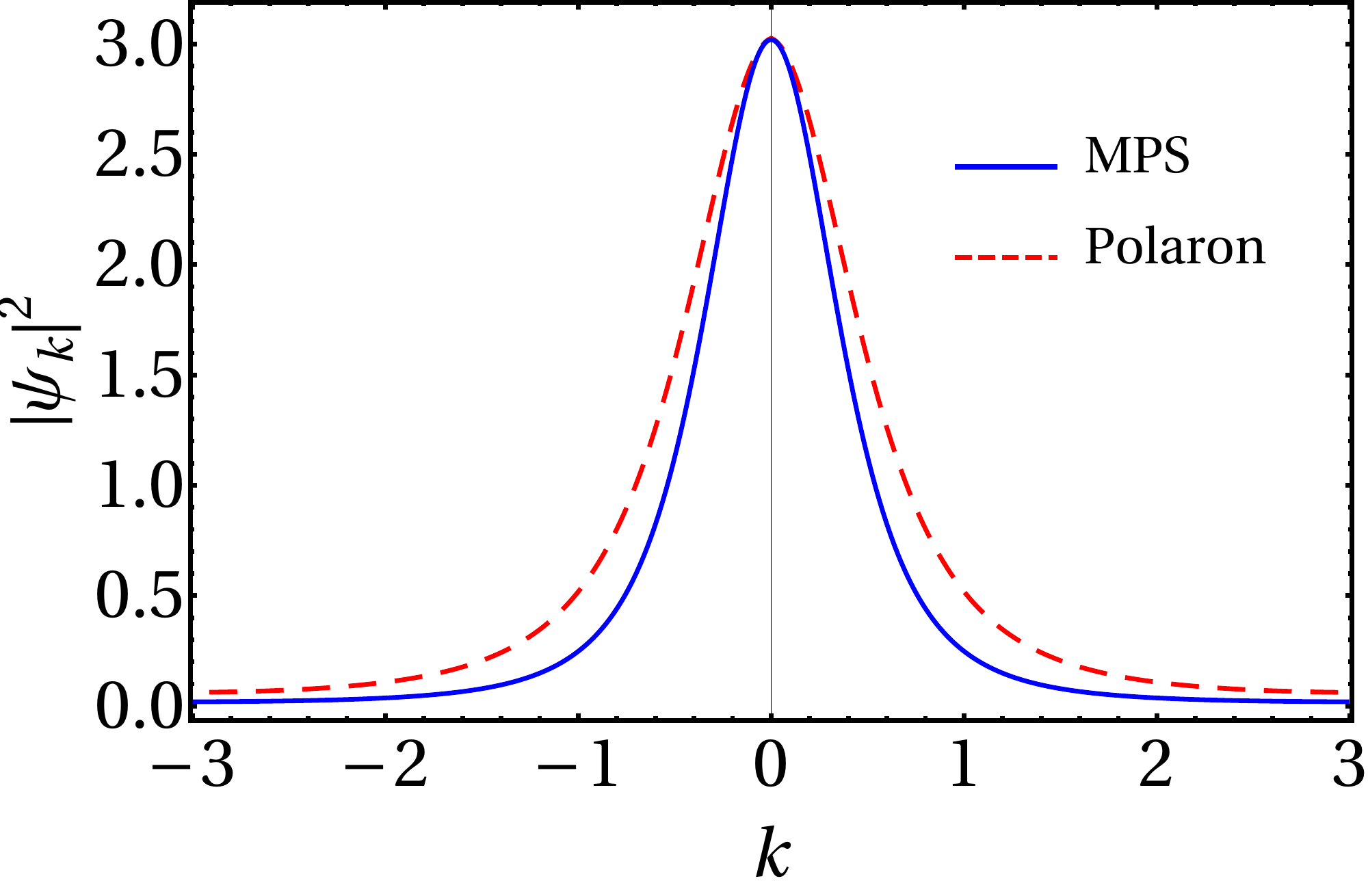}
\caption{We compare the single-photon component of the ground state computed with MPS and the polaron ansatz. The parameters are those of Fig. 3 of the manuscript.}\label{fig:vector}
\end{figure}

We now compare the ground state wave-function given by the polaron against the numerical MPS calculations.   Using the polaron picture, the ground state can be written as:
\begin{equation}
| \text{GS} \rangle = U_p |0 ; {\bf 0} \rangle
\end{equation}
with (Eq. (3) of the main text)
\begin{equation}
\label{Up-app}
  U_p = e^{-\sigma_x \sum (f_k a_k^\dagger - f_k^* a_k)} \; ,
\end{equation}
Expanding the exponential we get,
\begin{align}
\label{expa}
| \Psi_{\rm gs} \rangle=
e^{-\sum_k |f_k|^2}
\Big ( & 1 - \sigma^+ \sum_k f_k a_k^\dagger 
\\ \nonumber
&+ \frac{1}{2}  
\big ( \sum_k f_k a_k^\dagger 
\big )^2 + ... \Big )
|0 ; {\bf 0} \rangle
\; .
\end{align}
Therefore, the ground state one-photon coefficients (second term in \eqref{expa}) are given by (up to a normalization) $f_k$.  These coefficients are compared in Fig.  \ref{fig:vector} for $g=0.2$. The agreement is rather good.

\begin{figure}[tbh!]
\includegraphics[scale=0.40]{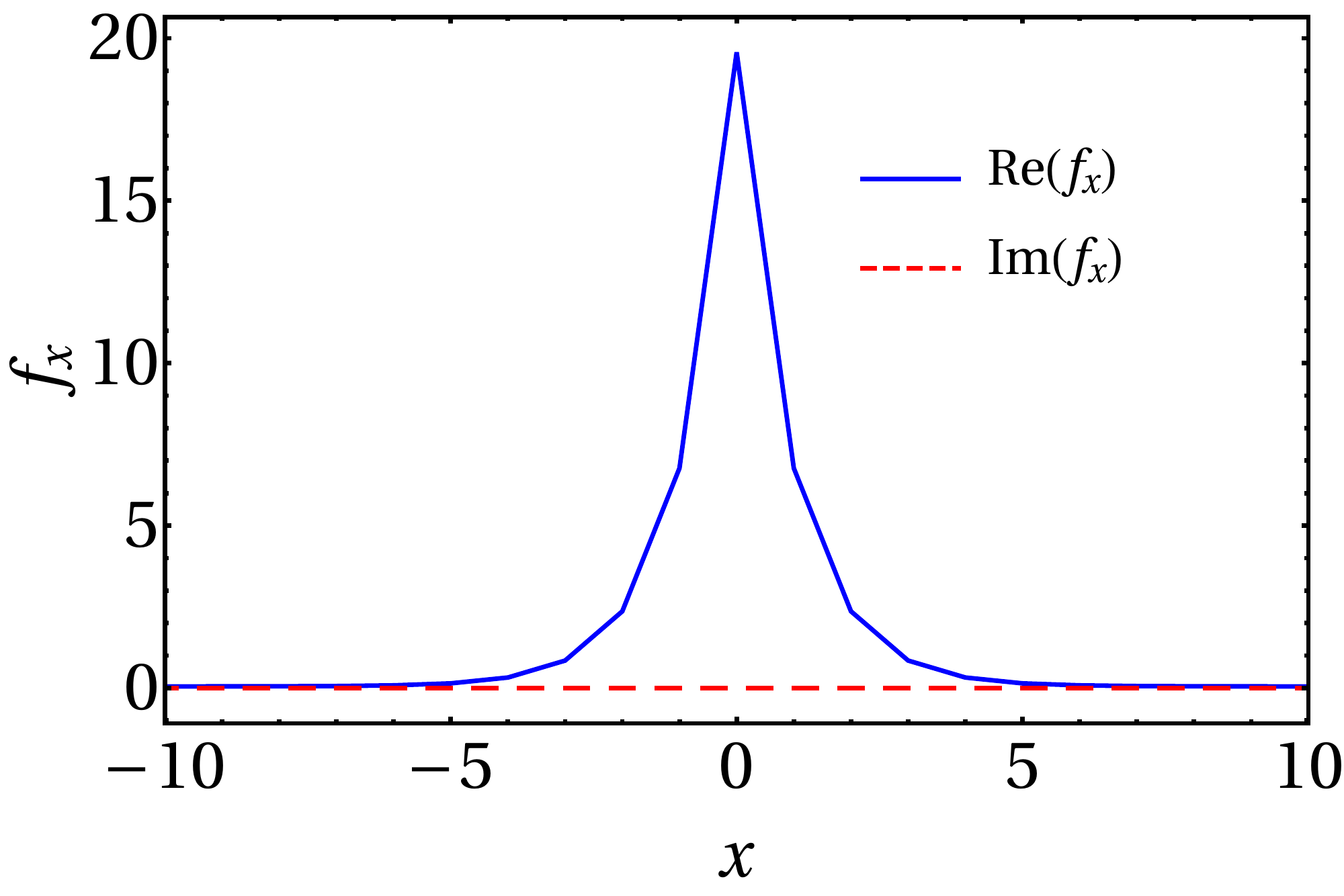}
\caption{Real (blue points) and imaginary (red points) parts of $f_x$, Eq. \eqref{eq:fx}. The coupling constantis $g=0.4$ and the rest are the same as in Fig. 3 of the manuscript.}\label{fig:fn}
\end{figure}

In the main text, we have assumed that the polaron trasnformation is virtually local. 
This means that far away from the 2LS the transformation is close to the identity. 
This makes sense, since the ground state is nontrivial only around the 2LS (Cf. Fig. \ref{Fig1}b)).
In order to verify this guess, we transform the polaron coefficients $f_k$ to real space:
\begin{equation}\label{eq:fx}
f_x = \frac{1}{\sqrt{N}} \sum _k e^{i 2 \pi k x / N} f_k \; .
\end{equation}
In Fig. \ref{fig:fn} we show that $f_x$ is different from zero only around the 2LS, demonstrating the local character for the polaron transformation in our case.

\section{Qubit dynamics in the quenching protocol}

\begin{figure}[tbh!]
\includegraphics[scale=0.25]{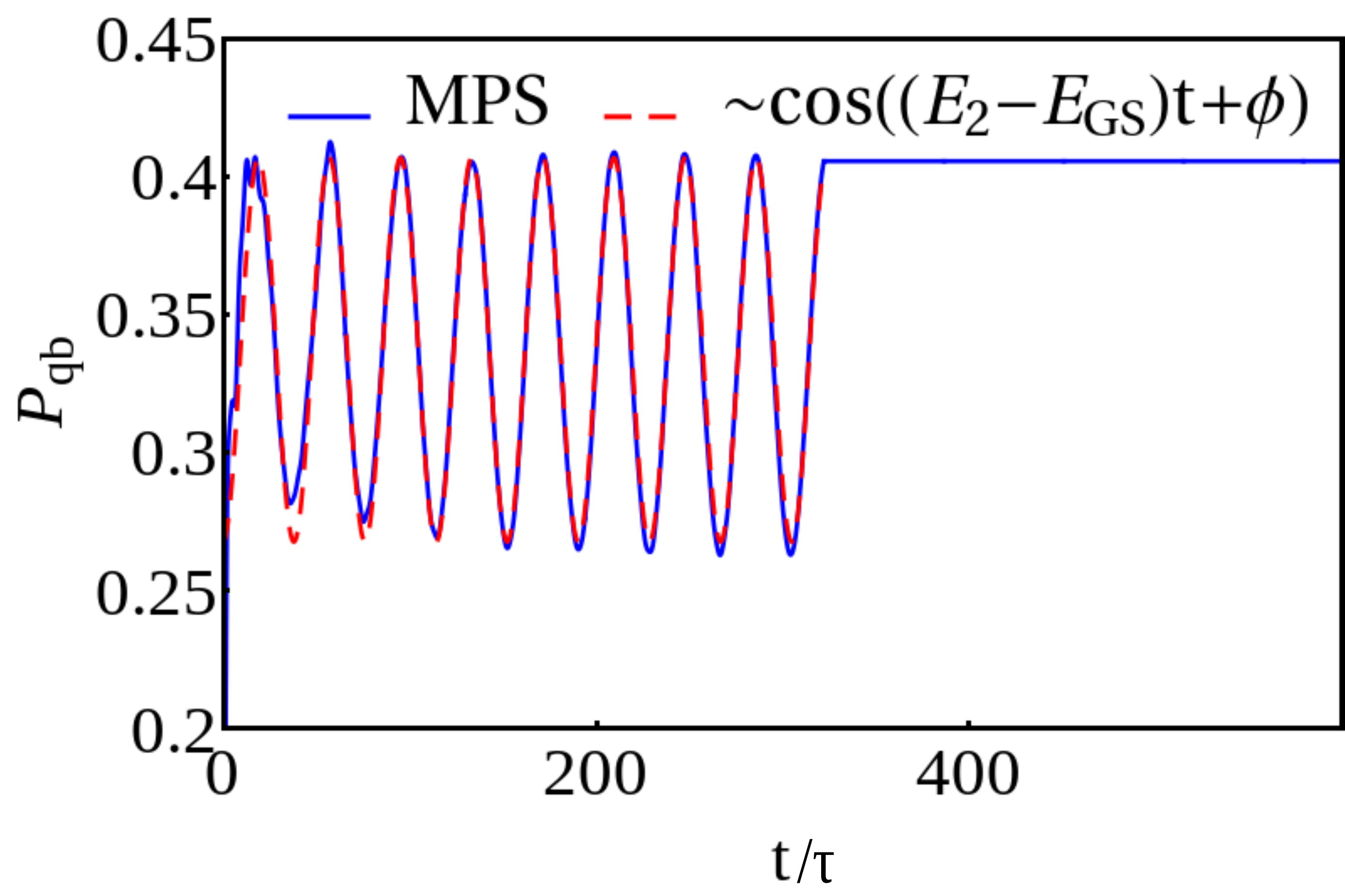}
\caption{Qubit dynamics for the quenching protocol. Same parameters as in Fig. 3 of the main text.} \label{fig:qubit}
\end{figure}

We prove here that $\ket{\Psi(t)}_p$ (Eq. (7) of the main text) explains the dynamics of the qubit. We plot the excited-state popuation $P_{\rm qb}(t) = \braket{\Psi (t) | \sigma^+ \sigma^- | \Psi (t)}$ in Fig. \ref{fig:qubit}. As seen, $P_{\rm qb}(t)$ oscillates with frequency $E_2-E_\text{GS}$, so it shows the interference of the amplitudes of $|\text{GS}\rangle$ and $|E_2\rangle$ (first two terms of Eq. (7) of the manuscript). It estabilizes after a transient  time of the order of the qubit relaxation time, $t\geq \tau$, and persist until the coupling is switched off at $t_\text{off}$, when $P_{qb}$ gets frozen.

%%%%%%%%%% bibliography

\bibliographystyle{apsrev4-1}
\bibliography{bib_vac}

\end{document}